\documentclass[Afour,sagev,times]{sagej}

\usepackage{moreverb,url}

\usepackage[colorlinks,bookmarksopen,bookmarksnumbered,citecolor=red,urlcolor=red]{hyperref}

\newcommand\BibTeX{{\rmfamily B\kern-.05em \textsc{i\kern-.025em b}\kern-.08em
T\kern-.1667em\lower.7ex\hbox{E}\kern-.125emX}}

\begin{document}

\runninghead{Statistical and Practical Considerations in Planning and Conduct of Dose Optimization Trials}

\title{Statistical and Practical Considerations in Planning and Conduct of Dose Optimization Trials }

\author{Ying Yuan\affilnum{1}, Heng Zhou\affilnum{2}, and Suyu Liu\affilnum{1}}

\affiliation{
\affilnum{1} Department of Biostatistics, University of Texas MD Anderson Cancer Center, Texas, USA \\
\affilnum{2} Biostatistics and Research Decision Sciences, Merck \& Co., Inc, Rahway,
New Jersey, USA \\
}

\corrauth{Ying Yuan, Department of Biostatistics, University of Texas MD Anderson Cancer Center, TX, 77030, USA.}

\email{yyuan@mdanderson.org}

\begin{abstract}
The US Food and Drug Administration launched Project Optimus with the aim of shifting the paradigm of dose-finding and selection towards identifying the optimal biological dose that offers the best balance between benefit and risk, rather than the maximum tolerated dose. However, achieving dose optimization is a challenging task that involves a variety of factors and is considerably more complicated than identifying the maximum tolerated dose, both in terms of design and implementation. This article provides a comprehensive review of various design strategies for dose optimization trials, including phase I/II and II/III designs, and highlights their respective advantages and disadvantages. Additionally, practical considerations for selecting an appropriate design and planning and executing the trial are discussed. The article also presents freely available software tools that can be utilized for designing and implementing dose optimization trials. The approaches and their implementation are illustrated through real-world examples.
\end{abstract}


\maketitle

\section{Background}

The conventional phase I dose-finding paradigm was developed during the era of cytotoxic therapies with the primary goal of identifying the maximum tolerated dose (MTD) based on dose-limiting toxicity (DLT). The underlying assumption of this more-is-better approach is that both efficacy and toxicity increase monotonically with the dose. However, concerns have been raised regarding the appropriateness of using this approach in the age of targeted therapies and immunotherapies. \cite{Ratain2014, Yan2018, Shah2021, Ratain2021} Many of these innovative therapies exhibit a shallow dose-response, meaning that the MTD may not be reached within a clinically effective dose range. Furthermore, efficacy may not increase monotonically with the dose, and often reaches a plateau after a certain level is reached. \cite{Sachs2016, Zirkelbach2022} Consequently, the MTD may provide minimal improvements in efficacy over a lower dose, while causing more adverse events. For these reasons, the focus of dose-finding and selection should be shifted from finding the MTD to the identification of the optimal biological dose (OBD).

In 2021, the U.S. Food and Drug Administration (FDA) Oncology Center of Excellence launched Project Optimus with the goal of reforming the dose optimization and selection paradigm in oncology drug development \cite{FDA2022}. To facilitate this shift, the FDA also released draft guidance titled ``Optimizing the Dosage of Human Prescription Drugs and Biological Products for the Treatment of Oncologic Diseases" \cite{FDA2023}. Zirkelbach et al. \cite{Zirkelbach2022} and Shah et al. \cite{Shah2021} have offered valuable insights into the rationale, significance, and principles of dose optimization, along with a few drug approval examples from a regulatory agency perspective. Poor dose optimization have several negative consequences, such as failure to bring a drug to market, frequent dose modifications at the approved dose, and post-marketing requirements to further evaluate the dose. Shah et al. \cite{Shah2021} provide examples of approved drugs whose dose was modified for safety or tolerability after approval. 

Dose optimization trials are more complex than conventional MTD-finding trials. \cite{Yuan16book} The latter mainly focuses on DLT and is guided by a simple decision rule: if the observed data suggest that the DLT probability of the current dose is substantially greater (or lower) than the target DLT rate, we de-escalate (or escalate) the dose. In contrast, dose optimization trials are inherently multidimensional. By definition, they require the characterization and assessment of the benefit-risk of the doses, which involves various data, including toxicity and efficacy data, as well as pharmacokinetics (PK), pharmacodynamic (PD), and biomarker data. The decision of dose transition and selection must be based on the benefit-risk assessment of the doses. This increased dimensionality complicates the trial design, decision rule, and their implementation, and often requires larger sample sizes. As a result, it likely increases costs and prolongs the timeline for early phase drug development. Therefore, it is critically important to have efficient and novel statistical design strategies to address these challenges and meet the increasing regulatory requirements for oncology dose justification.

The aim of this article is to provide statistical and practical considerations related to the planning and execution of dose optimization trials. To facilitate understanding, we classified dose optimization trials conducted pre-market into two types: phase I/II dose optimization, where dose optimization is performed in phase I or/and II, and phase II/III dose optimization, where dose optimization is performed in phase II or/and III. In what follows, we discuss the methods, challenges, and practical considerations for designing and implementing both types of trials. We also provide real-world trial examples to illustrate key considerations. Finally, we conclude with a brief discussion.

\section{Phase I/II Dose Optimization}
\begin{figure}[htbp]
\centering
\includegraphics*[scale=0.3]{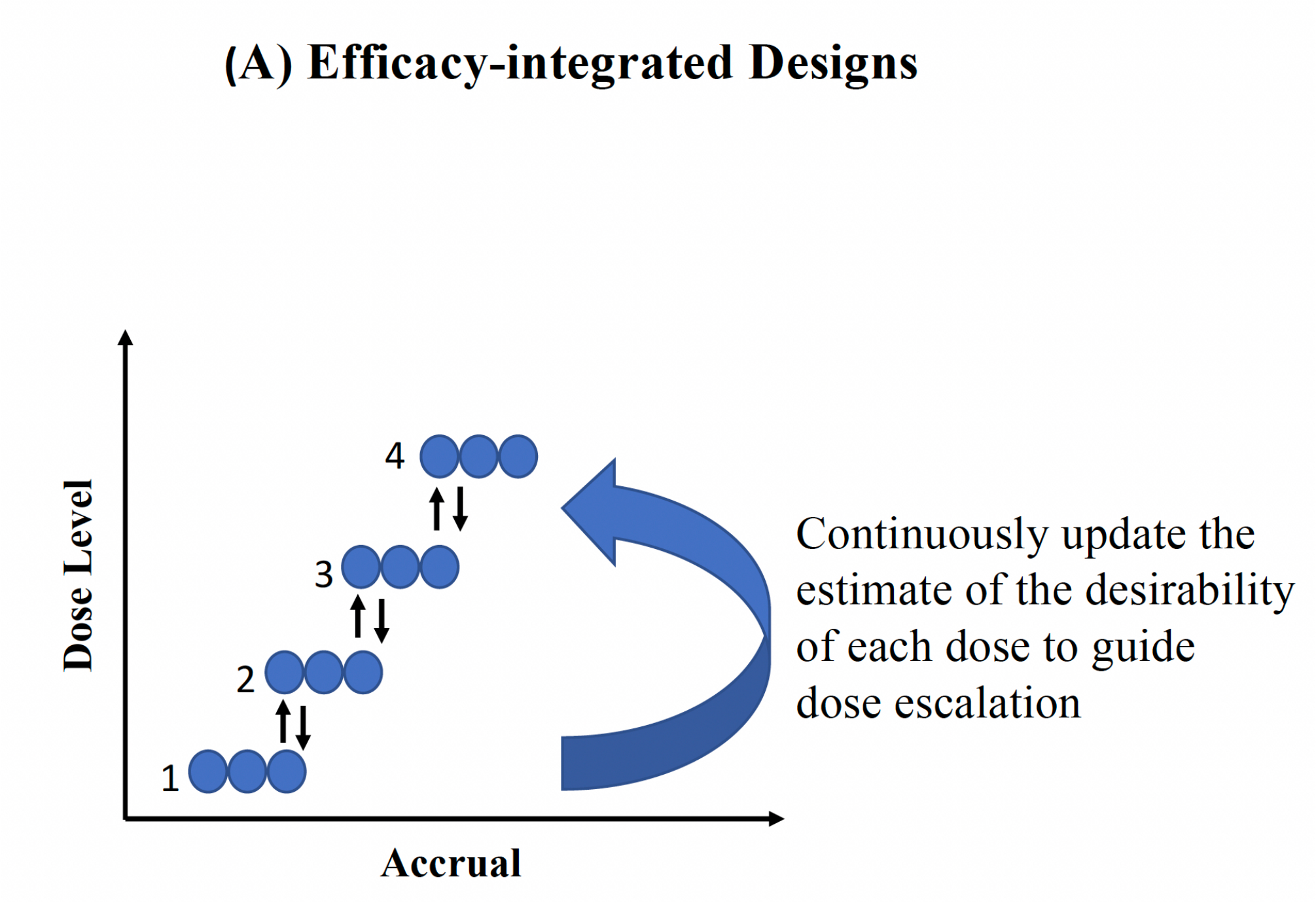}
\includegraphics*[scale=0.3]{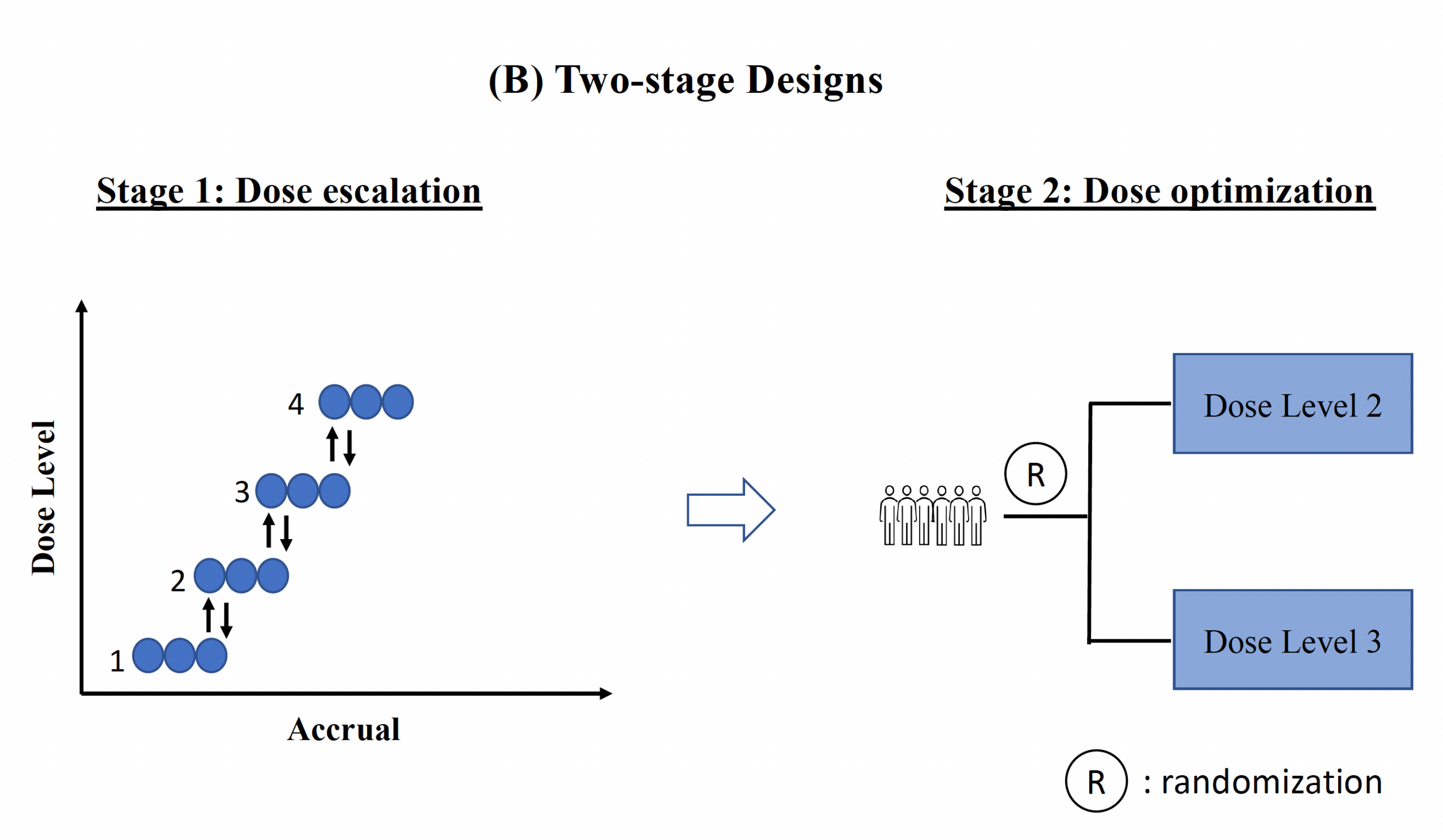}
\caption{(A) Efficacy-integrated designs, which achieve dose optimization by continuously updating the estimate of the benefit-risk tradeoff of each dose, based on the most recent data, to guide dose escalation, de-escalation, and selection. (B) Two-stage designs, where stage 1 dose escalation is performed to establish the MTD, and stage 2 conducts dose optimization often by randomization in multiple doses identified in stage 1.}\label{phase12}
\end{figure}

The topic of dose optimization has recently been propelled into the spotlight with the advent of Project Optimus, but the concept of finding the optimal dose based on a consideration of both risk and benefit is not a new one. \cite{Thall98, Quigley2001, Braun2002} Since Thall and Cook's seminal work on the EffTox design, \cite{Thall2004} a plethora of designs have been proposed for optimizing doses in phase I/II settings, \cite{Yin06,  Mandrekar07, Yuan09, Yuan11,  Wages14, Jin14, Zang2014, Liu16, Guo17, Riviere2018, Takeda2018, Lin2020} see the book of Yuan et al. \cite{Yuan16book} for a comprehensive review. To aid understanding of this vast and ever-increasing number of designs and provide a roadmap for future development, we categorize them as efficacy-integrated designs and two-stage designs (Fig \ref{phase12}). 

Efficacy-integrated phase I/II designs, also known as fully sequential phase I/II designs, directly target the OBD by performing dose escalation or de-escalation based on the benefit-risk tradeoff. The decision to escalate or de-escalate the dose is typically made by continuously updating the dose-toxicity and dose-efficacy model estimates based on interim data, in a similar fashion as the continuous reassessment method (CRM). \cite{Quigley1990} In contrast, two-stage phase I/II designs first identify the MTD through conventional DLT-based dose escalation, and then randomize patients among multiple doses to identify the OBD based on risk-benefit assessment. Examples of efficacy-integrated phase I/II designs include EffTox \cite{Thall2004} and BOIN12, \cite{Lin2020} while examples of two-stage phase I/II designs include the design by Hoering et al., \cite {Hoering2011} U-BOIN, \cite{Zhou2019} and DROID \cite{Guo2023}. It is important to note that the purpose of differentiating the two classes is to aid understanding and not to provide a definitive definition. A design may incorporate features of both approaches. For instance, a design may use a benefit-risk tradeoff to perform dose escalation, and meanwhile including randomization. \cite{Liu2018, Thall2023}

In the following sections, we first describe some essential elements of phase I/II designs, including efficacy and toxicity endpoints and benefit-risk tradeoff, and then provide a more detailed description of efficacy-integrated and two-stage design strategies.

\subsection{Efficacy and toxicity endpoints}

Dose optimization involves assessing the risks and benefits of a new drug, which requires specifying efficacy and toxicity endpoints to characterize its potential outcomes. An example of a toxicity endpoint is DLT, which is often defined as a grade 3 or higher adverse event that occurs within the first treatment cycle according to the Common Terminology Criteria for Adverse Events. However, DLTs may not be sufficient to fully characterize the safety and tolerability of many novel targeted drugs. These drugs often cause low-grade toxicity but rarely result in DLTs. In such cases, more comprehensive toxicity endpoints that account for low-grade or cumulative toxicity, such as ordinal toxicity endpoint \cite{Van2011}, total toxicity burden \cite{Bekele2004, Lee2012, Ezzalfani2013, Chen2010, Yuan2007, Mu2019} or equivalent toxicity score \cite{Yuan2007, Mu2019}, may be preferred. To quantify the safety profile of the drug when cumulative toxicity is expected, the dose tolerability rate over multiple cycles can also be used. \cite{Du2019}  The resulting toxicity endpoint can be binary, categorical, semi-continuous, or continuous. \cite{Mu2019}  Each type of toxicity endpoint has its advantages and limitations and should be chosen carefully with both clinical and statistical input. Alternatively, multiple constraints can be used to control different grades of toxicity at prespecified levels. \cite{Lee2009, Lin2011} 

Efficacy endpoints should be chosen based on clinical and logistic considerations. For early phase trials, objective response rate (ORR) along with the duration of response, measured by Response Evaluation Criteria in Solid Tumor (RECIST) guideline, are commonly used. However, the response often takes several cycles to be ascertained, which causes major logistic difficulties for making real-time dose assignment decisions. In this case, intermediate short-term endpoints, such as PD endpoints or target receptor occupancy, may be used instead. Alternatively, new designs are available to handle delayed responses. \cite{Jin14, Takeda2020, Zhou2022} A potential issue with using short-term endpoints is that they may not be a reliable surrogate of clinical endpoints. To address this, one approach is to use short-term endpoints to guide real-time dose assignment decisions during the trial, and then use the long-term clinical endpoint at the end of the trial to identify the OBD. \cite{Guo2023} In some situations, more than one efficacy endpoint can be used to capture the multifaceted effects of the drug and improve design efficacy. It is more convincing if different pieces of evidence, such as PK, PD, and efficacy, point to the same dose. For instance, Agrawal et al considered receptor occupancy, PK parameter, and tumor shrinkage jointly for nivolumab dose selection. \cite{Agrawal2016} In some trials, patient-reported outcomes, such as quality of life, are an important consideration for determining the OBD.

\subsection{Benefit-risk tradeoff}
The goal of dose optimization is to find the dose that achieves the optimal balance between benefit and risk. However, in current practice, the benefit-risk tradeoff (BRT) is rarely explicitly defined or used to guide dose optimization and selection. We believe that explicitly defining the BRT, tailored to the trial's objectives and characteristics, is advantageous and should be more widely adopted. By doing so, investigators can evaluate and fine-tune the design's operating characteristics, resulting in more efficient dose optimization.

A straightforward method to define the BRT is by considering the tradeoff between the probability of efficacy and toxicity of a dose. \cite{Liu16} For example, BRT $=\pi_E-w\pi_T$, where $\pi_E$ and $\pi_T$ are the probability of efficacy and toxicity of a dose, respectively; and $w>0$ represents the penalty for an increase in the toxicity rate. If two doses have similar $\pi_E$  but different $\pi_T$, the dose with the lower $\pi_T$ will have a higher BRT and will be preferred. Nonetheless, the BRT can also take more complex forms, such as nonlinear functions of $\pi_E$  and $\pi_T$, as illustrated in the EffTox design. \cite{Thall2004}

A more versatile and broadly applicable method of defining the BRT is through utility, \cite{Houede2010, Guo17, Murray2017, Liu2018, Lin2020} where clinicians are asked to provide utility scores for each potential patient-level outcome, thus quantifying the desirability of doses. For instance, with binary toxicity and efficacy endpoints, four potential outcomes exist for each patient: (efficacy, toxicity) = (yes, no), (yes, yes), (no, no), and (no, yes). We assign a score of 100 to the most desirable outcome (yes, no) and a score of 0 to the least desirable outcome (no, yes). For the remaining two outcomes, we elicit the scores from clinicians; for example, a score of 60 and 40 may be assigned to (yes, yes) and (no, no), respectively, see Table \ref{utilityscore}.  The BRT of a dose is then defined as the average of the scores, weighted by the probability of each potential outcome. That is, ${\rm BRT} = \sum_{k=1}^K \pi_k u_k$, where $u_k$ is the utility score assigned to the $k$th potential outcome, and $\pi_k$ is the probablity of observing the $k$th potential outcome, for $k=1, \cdots, K$. For instance, consider a dose with the probabilities of observing (no toxicity, efficacy), (no toxicity, no efficacy), (toxicity, efficacy), and (toxicity, no efficacy) being 0.5, 0.15, 0.25, and 0.1, respectively. The desirability of this dose is calculated as $0.5\times 100 + 0.15\times 40 + 0.25\times 60 + 0.1\times 0 = 71$. As a result, a dose with a higher probability to produce favorable outcomes will have a higher desirability. 

The utility-based BRT is often a better option than the probability-based BRT discussed earlier since clinicians typically have a better grasp of the relative desirability of patient outcomes than probabilities. Furthermore, research has shown that the utility approach encompasses the efficacy-and-toxicity-probability-based BRT as a special case. \cite{Zhou2019, Lin2020, Schipper2014SIM}

The use of a BRT to guide dose optimization may raise some concerns. The first concern is that BRT is subjective. However, we consider the need for elicitation from trial clinicians as a strength. This process encourages clinicians to carefully consider the benefit-risk tradeoff and design the study accordingly, reducing subjectivity and variability. Leaving the decision process unspecified actually leads to more subjectivity and variability. Additionally, specifying the BRT enables the evaluation of the design's operating characteristics through simulation, enhancing understanding and providing opportunities to improve the design before the trial begins. Lastly, research indicates that many designs are quite robust to the specification of BRT. To reduce the subjectivity of BRT elicitation, open communication and active efforts among stakeholders to reach a consensus are recommended. Seeking regulatory input early on during the design stage is also essential.

The second common concern regarding the use of BRT for dose optimization is that the benefit and risk of a treatment are multi-dimensional, making it challenging to capture all aspects with a single BRT. However, it is important to note that the BRT is primarily defined to facilitate and enhance the efficiency of the adaptive decision-making process for dose assignment and evaluation of the trial design's operating characteristics. Ultimately, the final decision for dose selection, particularly at the end of the trial, should be based on both the design recommendation, which is determined by the prespecified BRT, and the totality of the evidence.

\subsection{Efficacy-integrated designs}

The efficacy-integrated design is characterized by its use of toxicity and efficacy, which are combined simultaneously through the use of BRT, to guide dose escalation and de-escalation, and ultimately determine the OBD (Fig \ref{phase12}(A)). This approach assumes a statistical model to update the dose-toxicity and dose-efficacy relationship based on interim data. This information is then utilized to make adaptive dose assignment decisions that prioritize doses with high BRT for treating the next cohort of patients. The efficacy-integrated design operates similarly to the CRM, but uses the estimate of BRT rather than DLT to make dose transition decisions. For ease of explanation, the terms BRT and desirability are used interchangeably. Depending on the method and level of complexity of implementation, efficacy-integrated designs can be further differentiated into model-based designs and model-assisted designs.

\subsubsection{Model-based designs}
The model-based design assumes a statistical model to depict the dose-toxicity and dose-efficacy curves, which are used to guide dose transition. Examples of model-based designs include EffTox \cite{Thall2004} and late-onset (LO)-EffTox designs. \cite{Jin14} EffTox assumes the following logistic marginal dose-toxicity and dose-efficacy models: ${\rm logit}(\pi_T | x) = \gamma_0+\gamma_1 x$ and ${\rm logit}(\pi_E | x) = \beta_0+\beta_1 x + \beta_2 x^2$, respectively, where $x$ is a standardized dosage. Given the marginal models, the Gumbel-Morgenstern copula is further used to model the joint distribution of (toxicity, efficacy) = $(y_T,y_E)$ as follows: $f(y_T, y_E |x) = \pi_E^{y_E} (1-\pi_E)^{1-y_E} \pi_T^{y_T} (1-\pi_T)^{1-y_T} + (-1)^{y_T+y_E} \pi_E (1-\pi_E)\pi_T (1-\pi_T)\frac{e^\psi-1}{e^\psi+1}$, where $\psi$ is a parameter presenting the correlation between $y_T$ and $y_E$. During the trial, the dose-toxicity and dose-efficacy models are fitted based on the interim data, and the next cohort of patients is assigned to the dose with the highest estimate of BRT. Despite their desirable statistical properties, the use of model-based designs has been hindered by the requirement for complicated model estimation after each cohort and the influence of model misspecification.

\subsubsection{Model-assisted designs}
Model-assisted designs have been developed to address the limitations of model-based designs. \cite{Yuan2019JCO, Yuan2022book} Unlike model-based approaches, model-assisted approaches use simple models, such as binomial or multinomial models, at each dose, without assuming any specific shape for the dose-toxicity or dose-efficacy curves. As a result, the decision rule can be derived and tabulated before the trial onset. During the trial, users can simply consult the decision table to make dose assignment decisions. The book by Yuan et al. \cite{Yuan2022book} provides a comprehensive review of model-assisted designs. Due to their simplicity and desirable operating characteristics, model-assisted designs have gained popularity in recent years. For instance, the Bayesian optimal interval (BOIN) design, \cite{Liu2014} a model-assisted design, received the fit-for-purpose designation from the FDA as a tool for dose-finding in oncology. \cite{FDA2022FFP}

\begin{figure}[htbp]
\centering
\includegraphics*[scale=0.65]{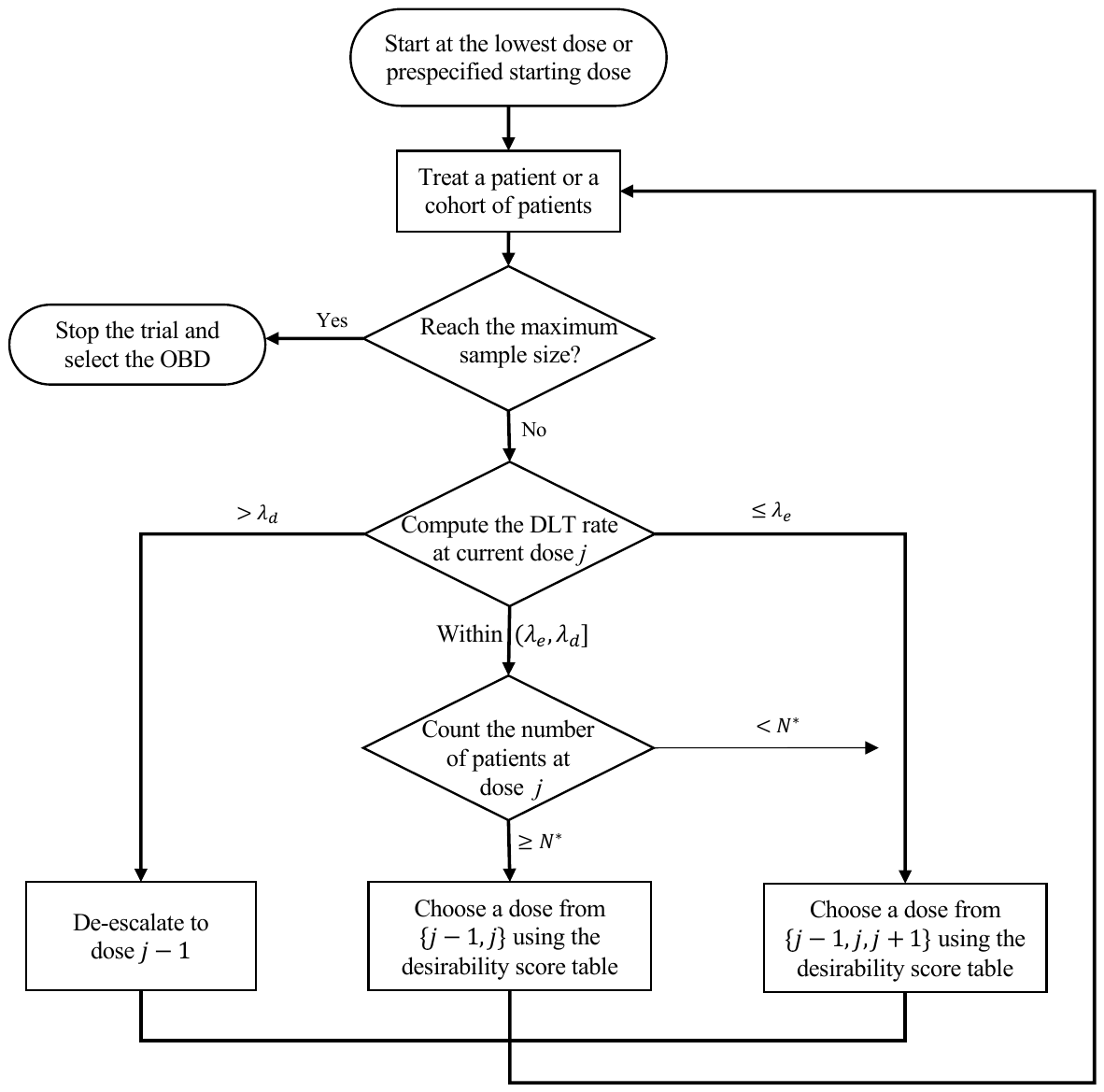}
\begin{minipage}{0.5\columnwidth}
\begin{tabular}{ccccccccc}
\hline
Target toxicity rate &   0.15 &        0.2 &          0.25 &        0.3 &      0.35 &        0.4 \\
\hline
$\lambda_e$ &       0.118 &      0.157 &      0.197 &         0.236 &         0.276    & 0.316\\

$\lambda_d$ &      0.179 &       0.238 &       0.298 &       0.358 &       0.419  & 0.479  \\
\hline
\end{tabular}  
\end{minipage}
\caption{The schema of the BOIN12 design, where $(\lambda_e, \lambda_d)$ are a pair of optimized dose escalation and de-escalation boundaries, and N* is a prespecified sample size cutoff (e.g., $N^*=6$). The desirability score table is provided in Table \ref{tab.RDS}.}\label{BOIN12}
\end{figure}

A number of model-assisted designs have been proposed for dose optimization. \cite{Takeda2018, Lin2020, Lin2017, Li2017, Shi2021} We here use BOIN12 to illustrate this approach. BOIN12 uses the utility to measure the toxicity-efficacy tradeoff and models it using a simple beta-binomial model based on pseudolikelihood methodology.  Based on interim toxicity and efficacy data, the BOIN12 design adaptively assigns patients to the dose with the highest estimated desirability. The dose-finding rule of BOIN12 is depicted in Fig \ref{BOIN12}. A key feature of this design is that dose desirability can be pre-tabulated and included in the trial protocol prior to the trial's start (see Table \ref{tab.RDS}), making implementation simple. During the trial, determining the desirability of a dose is straightforward: count the number of patients treated at a given dose, the number who experienced toxicity, and the number who experienced efficacy. This information is used to look up the dose's rank-based desirability score (RDS) in Table \ref{tab.RDS}. The next cohort of patients is then assigned to the dose with the highest RDS value. For instance, consider a scenario in which a trial has treated 3, 6, and 3 patients at the first three doses, respectively, and has observed toxicity and efficacy outcomes of (0, 1, 2) and (0, 3, 1), respectively. The current dose being administered is d = 2. In accordance with the dose-finding rule, we compare the observed toxicity rate $\hat{\pi}_T= 0.167$ to the escalation boundary $\lambda_e = 0.276$. Since $\hat{\pi}_T$ is less than the escalation boundary $\lambda_e$, we consult Table \ref{tab.RDS}, which provides the RDS of each dose as 13, 23, and 11, respectively. As dose level 2 has the highest RDS, the decision is to continue administering the current dose to the next cohort of patients. 

It is worth noting that Table \ref{tab.RDS} assumes a cohort size of 3. To account for the possibility that the number of evaluable patients may not be a multiple of 3, a more comprehensive decision table can be generated using the software described later, which includes every possible number of patients up to the maximum number of patients that can be treated at a dose. The BOIN12 has been shown to have desirable operating characteristics through an extensive simulation study, often outperforming more complex model-based phase I/II designs, such as the EffTox design. \cite{Lin2020}

\subsection{Two-stage designs}
The two-stage design approach takes a staged approach to dose optimization. \cite{Hoering2011, Pan2014, Zhou2019, Guo2023} As illustrated in Fig \ref{phase12} (B), in stage 1, dose escalation is performed to establish the MTD. At the end of stage 1, typically, the MTD and one or two lower doses that demonstrate appropriate anti-tumor activities and PK/PD characteristics are selected and proceed to stage 2 for dose optimization often via randomization. Stage 1 often uses conventional MTD-targeted dose-escalation methods, such as model-based CRM or model-assisted BOIN. Thus, the key question for this approach is how to design stage 2, especially in terms of sample size determination.	
	
Yang et al. (2023) proposed the MERIT (Multiple-dosE RandomIzed Trial design for dose optimization based on toxicity and efficacy) design to provide a systematic approach to determining the sample size for stage II randomization. \cite{Yang2023} As in practice the final selection of the OBD involves both statistical and non-statistical considerations, it is of limited value to control the statistical properties of the design solely in terms of the OBD selection. Thus, MERIT focuses on controlling the statistical properties, such as type I error and power, for identifying doses that are admissible to be the OBD.  The OBD can only be selected from the admissible doses. To define the admissible doses, let $\phi_{T, 0}$ denote the null toxicity rate that is considered high and unacceptable, $\phi_{T, 1}$ denote the alternative toxicity rate that is deemed acceptable. Similarly, let $\phi_{E, 0}$ and $\phi_{E, 1}$ denote the null and alternative efficacy rates that are deemed unacceptable and acceptable, respectively. A dose is considered OBD admissible if its toxicity rate  $\pi_T \le \phi_{T, 1}$ and its efficacy rate $\pi_E \ge \phi_{E, 1}$.  

MERIT considers a null hypothesis ($H_0$): none of the doses is OBD admissible, versus an alternative hypothesis ($H_1$): at least one dose is OBD admissible. The design derives the minimal sample size, along with decision boundaries, that satisfies a pre-specified requirement on generalized type I error and power. The generalized type I error and power is a modification of standard type I error and power to accommodate the unique features of multiple-arm dose optimization, see Yang et al. (2023) for details. \cite{Yang2023} 

Table \ref{tab.MERIT} shows the optimal sample size for a randomized trial with 2 doses, as well as the decision rule to determine if a dose is OBD admissible.
For example, suppose $(\phi_{T, 0}, \phi_{T, 1})=(0.4, 0.2)$ and  $(\phi_{E, 0}, \phi_{E, 1})=(0.1, 0.3)$, to achieve the (generalized) power of 70\% and maintain the (generalized) type I error of 0.2, we will need to randomize $n=24$ per dose. After the completion of the trial, a dose is considered as OBD admissible (meaning it can be chosen as the OBD based on the totality of evidence) if the number of efficacy $\ge m_E=5$ and the number of toxicity $\le m_T=7$.


Regarding the method of randomization, equal randomization is the most commonly used approach due to its ease of implementation and unbiased comparison between doses. Although response-adaptive randomization may seem attractive as it allows for more patients to be allocated to a more desirable dose, it often provides little benefit for multiple-dose randomization trials with small sample sizes. In fact, accrual may be nearly complete before the data start to skew the randomization towards better doses. Additionally, response-adaptive randomization is more logistically challenging and increases the likelihood of unbalanced patient characteristic distribution across arms, which can lead to biased estimates. \cite{Wathen2017} Equal randomization combined with safety and futility monitoring, such as using Bayesian optimal phase 2 (BOP2) design, \cite{Zhou2017BOP2} is often effective and allows for early stopping of overly toxic and futile doses during the trial.

The approach of randomizing patients among multiple doses for optimization is commonly used in non-oncology drug development and is referred to as dose-ranging. However, well-established dose-ranging methods in non-oncology therapeutic areas, such as the Multiple Comparison Procedure – Modelling (MCP-Mod) method, \cite{Bretz2005} are rarely used in oncology due to the unique characteristics and challenges of cancer drug development. \cite{Guo2023} 
To address this issue, Guo and Yuan (2023) developed an oncology-specific dose-ranging design referred to as DROID,  by combining the mature framework of non-oncology dose-ranging with oncology dose finding. \cite{Guo2023}


\subsection{Design choice}
The efficacy-integrated and two-stage strategies each have their own advantages and disadvantages, making them suitable for different scenarios. The two-stage approach is well-aligned with conventional develop-by-stage practices, and can accommodate different populations for the dose escalation and randomization stages. However, a potential drawback of the two-stage design is that the true optimal dose may be incorrectly excluded when transitioning from stage I to II due to unreliable toxicity and efficacy estimates based on a small stage I sample size. This issue can be partially addressed by backfilling patients during the dose escalation stage to obtain more data and increase the reliability of dose selection. However, this approach may still be limited by a small sample size. In addition, the two-stage approach generally requires larger sample sizes than the efficacy-integrated approach.

In contrast, the efficacy-integrated approach continuously learns the toxicity and efficacy profile of all doses throughout the trial, making it more efficient to identify the optimal dose and requiring smaller sample sizes. One limitation of this approach is that it requires efficacy and toxicity endpoints to be quickly observable enough to make adaptive decisions. However, methods such as TITE-BOIN12 have been proposed to address this limitation and facilitate real-time decision-making in the presence of pending toxicity or efficacy data. Additionally, the efficacy-integrated approach requires that the population used for dose optimization is comparable to that for subsequent phase IIb or III trials, which could be challenging when the target population is not clear. In this case, after phase I/II dose finding, we may conduct cohort expansion (e.g., basket trials) in potential target populations to confirm the OBD and establish the target population before proceeding to phase III trials. This strategy is also applicable to the two-stage approach.

The efficacy-integrated and two-stage approaches demand different sample sizes. For the two-stage approach, the recommended sample size is 6$\times J$ for the dose escalation portion,\cite{Yuan2022book} where $J$ is the number of doses under investigation, and is 20-40 patients per dose arm for the randomization portion to achieve reasonable power and type I error. \cite{Yang2023} For the efficacy-integrated approach, based on our experience,  a sample size of  6$\times J$ to 9$\times J$ generally yields reasonable operating characteristics. \cite{Yuan2022book} For example, given 4 doses under investigation, a reasonable sample size for the two-stage design is between 64 to 104 (assuming two doses selected for randomization), and that for the efficacy-integrated approach is between 24 to 36. It is important to note that these are rules of thumb. Given a specific trial, the sample size should be validated and calibrated using simulation to ensure reasonable operating characteristics. 

The efficacy-integrated and two-stage design strategies can be combined to achieve more efficient dose optimization, and they are not mutually exclusive. For instance, a trial can begin with an efficacy-integrated design (e.g., BOIN12) to optimize the initial dose efficiently, and then progress to the second stage with multiple-dose randomization to refine optimization using the MERIT design. In the generalized phase I/II design, a third randomized stage is added to further optimize the dose based on long-term endpoints. \cite{Thall2023}

\section{Phase II/III Dose Optimization}
A phase II/III design offers an alternative strategy for dose optimization. This design type encompasses a broad range of designs and can serve various purposes, including treatment selection, population selection, and endpoint selection,\cite{Stallard2011, Kunz2015}  and expediting the drug development process for accelerated approval. \cite{FDA2023AA}  Here, we focus on the phase II/III design for the purpose of dose optimization. 

In this context, the phase II component involves the random assignment of patients to multiple doses, with or without a control, to evaluate benefits and risks of each dose. The doses are typically selected based on factors such as toxicity, PK/PD, and preliminary efficacy data collected in the phase I dose escalation study, which should demonstrate reasonable safety and anti-tumor activity. At the end of phase II, an interim analysis is performed to determine the optimal dose that produces the most favorable benefit-risk trade-off for further investigation in the phase III component of the trial. The goal of phase III is to confirm the efficacy of the selected optimal dose with a randomized concurrent control or historical control. 

\subsection{Types of phase II/III designs}
\begin{figure}[htbp]
\centering
\includegraphics*[scale=0.8]{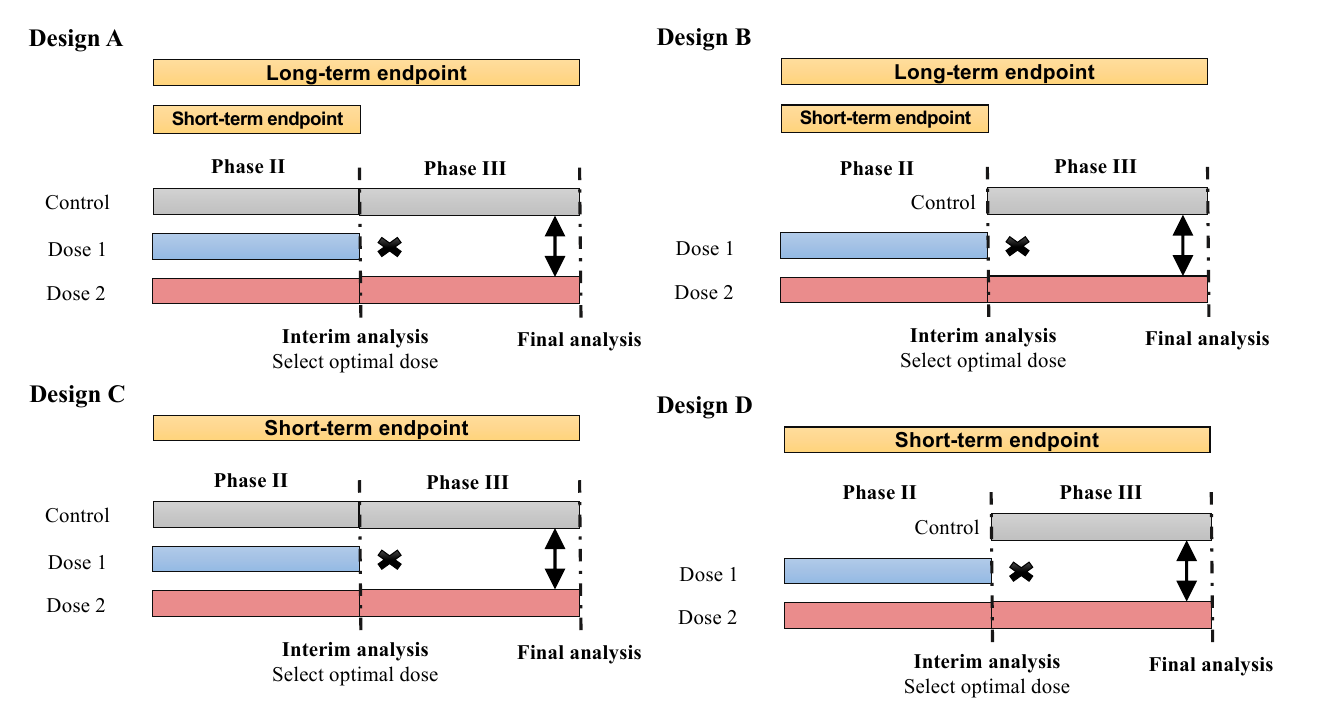}
\caption{Four types of phase II/III trial designs, varying in whether the concurrent control is included and the type of endpoints used in phase II and III.}\label{phase23}
\end{figure}

Depending on whether the concurrent control is included and the type of endpoints used in phase II and III, Jiang and Yuan \cite{Jiang2022} distinguish four forms of phase II/III dose-optimization designs (Fig \ref{phase23}) that are suitable for different clinical settings. 

Design A incorporates a concurrent control in both stages and employs a short-term binary endpoint (e.g., ORR) in phase II to identify the optimal dose. In phase III, a long-term time-to-event endpoint (e.g., progression-free survival (PFS) or overall survival) is used to assess the treatment's therapeutic effect. The use of a short-term endpoint in phase II allows for a prompt selection of the optimal dose to progress to phase III. Although not depicted in the schema, when appropriate, phase III may include an additional interim futility/superiority analysis akin to the standard group sequential design. An example of Design A is the HORIZON III trial for advanced metastatic colorectal cancer, \cite{Schmoll2012}which will be further elaborated in the Trial Examples section.

Design B is a modification of Design A that includes only the control in phase III. This can further reduce the sample size. However, Design B's drawback is the lack of concurrent control in phase II, making it difficult to combine phase II and III data and obtain an unbiased estimate of the treatment effect if there is a drift in the patient population or/and the treatment effect. Design B is a reasonable option when a drift is unlikely, for example when the accrual is fast such that the patient population is unlike to change, and the characteristics and performance of study centers remain stable over the trial period.  Design B was used in several clinical trials, such as a randomized multi-center trial of SM-88 in patients with metastatic pancreatic cancer. \cite{SM88}
	
Design C is similar to Design A but simpler because it employs the same short-term endpoint (e.g., ORR) for both phase II and phase III. This design is particularly useful in situations where demonstrating an effect on a long-term endpoint (e.g., survival or morbidity) requires lengthy and often large trials due to the disease's prolonged course, and the short-term endpoint is reasonably likely to predict clinical benefit on the long-term endpoint. The seamless phase II/III TNK-S2B trial of intravenous tenecteplase versus standard-dose intravenous alteplase for treating patients with acute ischemic stroke is an example of Design C. \cite{Levin2011} 

Design D is a simplified version of Design C that does not include a control. This design is appropriate when there is a particularly acute unmet medical need (e.g., a refractory or resistant patient population), and/or the tumor under treatment is rare. Designs C and D are useful for drug development that targets accelerated approval from the FDA, which often relies on a short-term surrogate or intermediate clinical endpoint such as response. A limitation of Design D, like Design B, is the lack of concurrent controls, which may result in a biased estimate of the treatment effect if there is a drift in the patient population.

In addition to considering short-term phase II and long-term phase III endpoints, additional endpoints can be employed for making adaptive decisions regarding the transition from phase II to III. One such approach is the 2-in-1 phase II/III designs, \cite{Chen2018CCT, Zhang2022CCT} which may provide additional flexibility for dose optimization.

\subsection{Operational versus inferential}
The phase II/III design can be categorized as either operational or inferential. In operational phase II/III designs, both phases are conducted under the same protocol to eliminate any gaps between them and reduce the overall trial cost. The data collected in phase II is not used in the phase III confirmatory analysis. Operational phase II/III designs are relatively simple to implement, and provide flexibility in dose selection and study design adjustments based on the results of the phase II portion while maintaining the integrity and reliability of the confirmatory phase III analysis.

Inferential phase II/III designs integrate both phase II and phase III data to evaluate the treatment effect. Because they incorporate additional phase II data, inferential phase II/III designs typically demand smaller sample sizes, shorten timelines of drug development, and exhibit greater statistical efficiency compared to operational phase II/III designs. However, for the same reason, they require careful consideration and specialized statistical methods to control the family-wise error rate, see Stallard and Todd \cite{Stallard2011} and Kunz et al. \cite{Kunz2015} for relevant methods. In addition, the implementation of inferential phase II/III designs is more logistically and operationally challenging compared to operational phase II-III designs, as detailed below.

\subsection{Practical considersions}
In operational phase II/III designs, two phases are conducted independently and the data collected in phase II is not used in the phase III confirmatory analysis. Therefore, the phase II and III portions can be conducted using standard considerations for phase II and III trials, respectively, resulting in little additional complexity beyond what is expected for each individual phase.

The use of inferential phase II/III designs for dose optimization presents more logistical and operational challenges. Due to the complexity of the design, it is crucial that sponsors and regulatory agencies engage in discussions about the trial design as early as possible in the development process. This allows for agencies to communicate their expectations and potentially leads to more efficient studies. Determining the doses to be studied in phase II/III trials requires knowledge of therapeutic properties, patient population heterogeneity, the need for additional dose exploration for a supplemental application, as well as communication between patients and providers. Similar considerations also apply to the selection of the optimal dose when stage 1 is complete.

For phase II/III trials to select the optimal dose, unblinded data access is necessary at the end of stage 1. However, this could compromise the trial integrity if not handled appropriately. The FDA guidance on adaptive designs recommends limiting access to comparative interim results to individuals with relevant expertise who are independent of the trial's conducting and managing personnel and who have a need to know. An Independent Data Monitoring Committee (IDMC) or an independent adaptation body should make the interim dose selection decision. Procedures must be in place to ensure that personnel responsible for preparing and reporting interim analysis results to the IDMC are physically and logistically separated from the trial's managing and conducting personnel. This requires planned procedures to maintain and verify confidentiality, as well as documentation of monitoring and adherence to operating procedures. To maximize the trial's integrity, investigators should prespecify design details, including the anticipated number and timing of interim analyses, criteria for dose selection, methods for controlling type I error and estimating treatment effects, and the data access plan.

\section{Software}
Designing dose optimization trials is more challenging than conventional dose-finding trials, and requires the use of more complicated statistical designs. It is critical to thoroughly evaluate and calibrate the operating characteristics before beginning the trial. Moreover, for certain designs, such as model-based designs, the trial conduct also requires real-time model fitting and calculation. Thus, easy-to-use software is key to the success of dose optimization trials. This is an area that requires immediate attention and development. The website \url{www.trialdesign.org} offers a dose optimization module that includes several dose optimization designs, such as BOIN12, TITE-BOIN12, MERIT, and DROID. Additionally, software to implement the EffTox design is available at from the software download website at The University of Texas MD Anderson Cancer Center.

\section{Trial Examples}

\subsection{Efficacy-integrated phase I/II dose optimization}
An efficacy-integrated model-based LO-EffTox design \cite{Jin14} was utilized to determine the optimal dose of sitravatinib in combination with a fixed dose of nivolumab for the treatment of clear cell renal cell carcinoma. \cite{Msaouel2022}  The trial had two primary endpoints: toxicity, defined as the time to DLT within 12 weeks of starting therapy, and early efficacy, defined as absence of progressive disease at 6 weeks using RECIST guideline by investigator assessment. Dose escalation/de-escalation was performed based on the benefit-risk trade-off constructed using marginal toxicity and efficacy probabilities. At the end of the trial, the 80 mg and 120 mg doses had almost the same estimated trade-off desirability scores; thus, additional criteria were used to compare the doses, including an evaluation of quality of life. The 120 mg dose was chosen as the optimal dose for sitravatinib and is currently being evaluated in ongoing phase II and III clinical trials for various malignancies. The implementation of this model-based phase I/II design has been logistically and resource-demanding. It requires a dedicated staff biostatistician to maintain frequent day-to-day communication between the clinical and data teams and perform real-time calculations to determine the dose assignment for the next patients. Tidwell et al. \cite{Tidwell2021} review this process and provide a summary of challenges and potential solutions, one of which is to use model-assisted designs such as BOIN12 as described next.

As an example, an optimal dose-finding trial of donor-derived CD5 CAR T cells in patients with relapsed or refractory T-Cell acute lymphoblastic leukemia was based on the BOIN12 design \cite{Pan2022}. The utility function shown in Table \ref{utilityscore} was used to measure the benefit-risk tradeoff of the treatment, and the RSD Table \ref{tab.RDS} was used to guide dose escalation. Patients were treated in cohorts of 3. Up to the time of reporting, a total of five patients who had CD7-negative relapse after CD7 CAR therapy were enrolled and received prior SCT donor-derived CD5 CAR T cells at an initial dose of $1\times10^6$ CAR T cells/kg. No DLT occurred, and all five patients achieved complete remission at day 30. It is important to note that because the first three patients all achieved complete remission, based on the BOIN12 design rule and RSD table, the trial continued to treat the second cohort at the dose of $1\times10^6$ CAR T cells/kg. However, if the standard dose escalation design were used, the dose would be increased to a higher level after the first three patients showed no DLT, which is unlikely to further improve efficacy but at the risk of more toxicity and a higher burden of manufacturing high levels of CAR T cells. This demonstrates the adverse effect of ignoring efficacy in conventional dose-finding designs and highlights the importance of performing dose optimization and the advantages of a dose optimization design.

\subsection{Two-stage phase I/II dose optimization}
Belantamab mafodotin, an antibody-drug conjugate targeting B-cell maturation antigen, was developed in a two-stage phase I/II study. In the DREAMM-1 first-in-human trial, \cite{Trudel2018} the dose escalation was performed based on the CRM design to explore doses ranging from 0.03 to 4.6 mg/kg. MTD was not reached. The 3.4 mg/kg dosage showed activity in the dose-expansion portion of DREAMM-1, but many patients experienced dose interruptions (71\%) and reductions (66\%). To improve tolerability, both 2.5 and 3.4 mg/kg doses were further evaluated in the DREAMM-2 trial in which patients were randomly assigned between the two dose arms. \cite{Lonial2020} Efficacy was similar between the 2.5 mg/kg cohort ($n=97$) with an ORR of 31\% (97.5\% CI, 20.8 to 42.6) and the 3.4 mg/kg cohort ($n=99$) with an ORR of 34\% (97.5\% CI, 23.9 to 46.0). There were fewer fatal AEs, serious AEs, dose interruptions, and dose reductions in patients receiving the 2.5 mg/kg IV. Exposure-response analysis showed a flat relationship, while a positive exposure-safety relationship was observed for keratopathy toxicity. Therefore, the 2.5 mg/kg IV was recommended, and the drug was granted accelerated approval. However, post-market commitment is required to optimize the dose due to ocular toxicity.

\subsection{Phase II/III dose optimization}
The HORIZON III trial compared the efficacy of cediranib with that of bevacizumab when used in combination with chemotherapy mFOLFOX6 for first-line treatment of advanced metastatic colorectal cancer. \cite{Schmoll2012} The trial employed a randomized, double-blind, inferential phase II/III design (Design A). During the phase II part, patients were randomly assigned 1:1:1 to receive cediranib 20 or 30mg per day or bevacizumab 5 mg/kg intravenous infusion every 14 days, each combined with 14-day treatment cycles of the regimen. An IDMC conducted end-of-phase II data analysis after 225 patients had three months of follow-up. The IDMC concluded that cediranib 20 mg met all predefined criteria for continuation. As a result, patients enrolled in the phase III part of the study were randomly assigned 1:1 to receive mFOLFOX6 with cediranib 20 mg or bevacizumab. All study personnel other than the IDMC remained blinded to the data until the trial ended. Patients who received cediranib 30 mg in the phase II part were unblinded and given the option to continue on open-label cediranib (20 or 30 mg per day). The primary analysis was planned for the primary endpoint PFS, which would occur after 850 progression events had occurred, based on all data from patients recruited into both the phase II and phase III parts of the study, excluding data from the cediranib dose discontinued at the end of phase II. Since PFS data from patients recruited into the phase II part of the study were used in the phase III analyses, and data from the phase II part were used to select the phase III dose, the method of Todd and Stallard (2005) \cite{Todd2005} was used to adjust the type I error for the primary analysis. The primary analysis showed that PFS had no significant difference between the arms. The estimated HR was 1.10 (95\% CI, 0.97 to 1.25), and the median PFS was 9.9 months for cediranib 20 mg and 10.3 months for bevacizumab. However, since the upper 95\% CI was beyond the predefined limit of 1.2, noninferiority of cediranib versus bevacizumab could not be concluded.

\section{Discussion}
We have reviewed design strategies and provided practical guidance on dose optimization trials. For phase I/II designs, we contrast efficacy-integrated and two-stage phase I/II design strategies and discussed their pros and cons and key considerations for trial implementation. For phase II/III designs, we discuss and compare different types of designs based on the type of endpoint, whether the control is included, and whether phase II data are combined with phase III data for the primary analysis (inferential or operational). 

In practice, the decision of whether to pursue a phase I/II or a phase II/III strategy should be made on a case-by-case basis, taking into account a variety of factors including clinical, statistical, logistic, and budgetary considerations. For instance, if a drug candidate has a fast readout of efficacy and PD parameters, an efficacy-integrated phase I/II design may be preferred to evaluate safety and efficacy simultaneously, starting early in the trial. On the other hand, if a drug candidate has been tested in a sufficient number of patients at various dose levels and there is a good understanding of its therapeutic window, a seamless phase II/III design may be a more attractive approach to selecting a dose from phase II and seamlessly bring it to the confirmatory phase III.

We have focused on dose optimization in the early phases (e.g., Phase I and II) of drug development, which is generally preferred to pre-market dose optimization. This approach increases the likelihood that the recommended dosage of the marketed product maximizes efficacy and minimizes toxicity, and avoids many issues associated with post-marketing dose optimization, such as the requirement for large sample sizes, long study durations, and difficulties in conducting the study as patients and investigators may be reluctant to be randomized to a dose of an approved product that differs from the approved dose. However, it is not uncommon that a dose has not been optimized at the time of marketing approval, and dose optimization studies are conducted after the drug has been approved. In such cases, the post-approval study evaluating two or more doses may be planned as non-inferiority trials. One may be concerned that performing dose optimization in the early phase will needlessly expose large numbers of patients to ineffective therapies and slow down drug development. \cite{Korn2022} Nevertheless, novel statistical designs, such as BOIN12 and EffTox, can stop the trial early when the drug demonstrates little activity, alleviating these concerns. Further research is warranted to develop and implement better study designs to maximize the benefit of dose optimization and deliver safe and effective treatments to patients.


%
%
%
%

\section{Declaration of conflicting interests}
The authors declared no potential conflicts of interest with respect to the research, authorship, and/or publication of this article.


\clearpage

\clearpage

\clearpage

\clearpage

\begin{table}
\caption{An example of utility for binary toxicity and efficacy endpoints.} \label{utilityscore}
\begin{tabular}{ccc}
\hline\hline
                 & Efficacy (Yes) & Efficacy (No) \\
                 \hline
Toxicity (Yes) & 60 & 0 \\
Toxicity (No) &100 & 40 \\
\hline\hline
\end{tabular}
\end{table}

\begin{table}
\caption{Rank-based desirability score (RDS) table for the BOIN12 design with the utility score 100 = (no toxicity, efficacy), 60 = (toxicity, efficacy), 40 = (no toxicity, no efficacy), and 0 = (toxicity, no efficacy).}\label{tab.RDS}
\begin{tabular}{cccc|ccccc}
\hline\hline
No.  & No.  & No. & RDS &  & No.  & No.  & No. & RDS\\
Pts & Tox & Eff &  &  & Pts & Tox & Eff & \\
\hline
0 & 0 & 0 & 24 &  & 6 & 1 & 1 & 10 \\
3 & 0 & 0 & 13 &  & 6 & 1 & 2 & 16 \\
3 & 0 & 1 & 22 &  & 6 & 1 & 3 & 23 \\
3 & 0 & 2 & 31 &  & 6 & 1 & 4 & 30 \\
3 & 0 & 3 & 38 &  & 6 & 1 & 5 & 36 \\
3 & 1 & 0 & 9 &  & 6 & 1 & 6 & 40 \\
3 & 1 & 1 & 17 &  & 6 & 2 & 0 & 2 \\
3 & 1 & 2 & 25 &  & 6 & 2 & 1 & 6 \\
3 & 1 & 3 & 33 &  & 6 & 2 & 2 & 12 \\
3 & 2 & 0 & 4 &  & 6 & 2 & 3 & 18 \\
3 & 2 & 1 & 11 &  & 6 & 2 & 4 & 26 \\
3 & 2 & 2 & 19 &  & 6 & 2 & 5 & 32 \\
3 & 2 & 3 & 29 &  & 6 & 2 & 6 & 37 \\
3 & $\ge$ 3 & Any & E &  & 6 & 3 & 0 & 1 \\
6 & 0 & 0 & 7 &  & 6 & 3 & 1 & 3 \\
6 & 0 & 1 & 14 &  & 6 & 3 & 2 & 7 \\
6 & 0 & 2 & 20 &  & 6 & 3 & 3 & 14 \\
6 & 0 & 3 & 27 &  & 6 & 3 & 4 & 20 \\
6 & 0 & 4 & 34 &  & 6 & 3 & 5 & 27 \\
6 & 0 & 5 & 39 &  & 6 & 3 & 6 & 34 \\
6 & 0 & 6 & 41 &  & 6 & $\ge$ 4 & Any & E \\
6 & 1 & 0 & 5 &  &  &  &  & \\
\hline\hline
\end{tabular}
Note: “Pts.,” “Tox.,” and “Eff.” denote patients, toxicity, and efficacy, respectively. “E” means that the dose should be eliminated, as it does not satisfy the safety and efficacy admissible criteria (i.e., not admissible due to high toxicity or low efficacy) with the upper toxicity limit of 0.35 and the lower efficacy limit of 0.25 and the probability cutoff of 0.9 for the admissibility. 
\end{table}

\begin{table}
\caption{Optimal design parameters for two doses when $(\phi_{T,1}, \phi_{T,0} )=(0.2,0.4)$.} \label{tab.MERIT}
\begin{tabular}{ccccccccccc}
\hline\hline
 &    & & \multicolumn{3}{c}{$\alpha = 0.1$} &&  \multicolumn{3}{c}{$\alpha = 0.2$} \\
\cline{4-6} \cline{8-10}
$\phi_{E,0}$ & $\phi_{E,1}$ & $\beta$ & $n$ & $m_T$ & $m_E$ && $n$ & $m_T$ & $m_E$ \\
\hline
0.1 & 0.3 & 0.6 & 25 & 6 & 5 && 18 & 5 & 4\\
 & & 0.7 & 33 &  8 & 6 && 24 & 7 & 5\\
 & & 0.8 & 39 & 11 & 8 && 30 & 8 & 5\\
 \hline
0.2 & 0.4 & 0.6 & 26 & 7 & 9 && 18 & 5 & 6\\
 & & 0.7 & 34 & 9 & 11 && 25 & 7 & 8\\
 & & 0.8 & 45 &  12 & 14 && 35 & 10 & 10\\
  \hline
0.3 & 0.5 & 0.6 & 28 & 7 & 12 && 19 & 5 & 8\\
 & & 0.7& 37 & 10 & 16 && 28 & 8 & 12\\
 & & 0.8& 44 & 12 & 18 && 34 & 10 & 14\\
  \hline
0.4 & 0.6 & 0.6 & 28 & 7 & 15 && 19 & 5 & 10\\
 & & 0.7 & 38 & 10 & 20 && 25 & 7 & 13\\
 & & 0.8 & 46 & 13 & 24 && 32 & 9 & 16\\
\hline\hline
\end{tabular}
Note: $\alpha$ and $\beta$ are pre-specified generalized type I error and generalized power, respectively.  $(n, m_T,m_E)$ are the optimal design parameters for sample size, and critical values for toxicity and efficacy responses, respectively. A dose is considered as OBD admissible (i.e., meaning it can be selected as the OBD based on the totality of evidence) if the number of efficacy $\ge m_E$ and the number of toxicity $\le m_T$. $\beta$ is the generalized power II defined in Yang et al. (2023) \cite{Yang2023}. 
\end{table}

\end{document}